\documentclass[12pt]{article}
\begin{document}

\title{\textbf{Generalized Intelligent States for Nonlinear Oscillators}}
\author{A. H. El Kinani$^{\dagger }$ and M. Daoud$^{\ddagger }$ \vspace{0.5cm} \\
$^{\dagger }$L.P.T, Physics Department, Faculty of sciences, \\
University Mohammed V, P.O.Box 1014 \\
Rabat, Morocco\vspace{1cm}\\
$^{\ddagger }$L.P.M.C, Physics Department, Faculty of sciences, \\
University Ibn Zohr, P.O.Box 28/S \\
Agadir, Morocco}
\date{}
\maketitle

\begin{abstract}
The construction of Generalized Intelligent States (GIS) for the $x^4$%
-anharmonic oscillator is presented. These GIS families are required to
minimize the Robertson-Schr\"odinger uncertainty relation. As a particular
case, we will get the so-called Gazeau-Klauder coherent states. The
properties of the latters are discussed in detail. Analytical representation
is also considered and its advantage is shown in obtaining the GIS in an
analytical way. Further extensions are finally proposed.
\end{abstract}

\newpage\

\section{Introduction and motivation}

The study of coherent states for a quantum mechanical system has received a
lot of attention $\left[ 1-4\right] $ and the definitions, applications,
generalizations of such states have been the subject of many papers. In the
recent years, they were discussed in connection with exactly solvable models
and non-linear algebras $\left[ 5-8\right] $ as well as deformed algebras $%
\left[ 9\right] $. They were also produced by using supersymmetric methods $%
\left[ 10-11\right] $. More recently, a new approach has been introduced
defining the set of coherent states, for a general quantum system, as
eigenstates of an annihilation operator which maintain all the properties
known for the standard harmonic oscillator $\left[ 12-14\right] $ (see also $%
\left[ 15\right] $ where an illustration of such construction is given for
infinite square well and P\"oschl-Teller potentials).

In this paper, we study the so-called generalized intelligent states for the
one-dimensional $x^4$-anharmonic oscillator. We recall that this quantum
system has been extensively studied since the early 1970 (for a review see $%
\left[ 16\right] $). Due mainly to its equivalence to one-dimensional $\phi
^4$ quantum field theory, it was hoped that a detailed study of this
simplified system would shed some light on the $\phi ^4$ theory in higher
dimensions. Research in this direction continues today. Indeed, recently A.
D. Speliotopoulos [$17$, and references therein] purposed a new method to
study the general structure of the Hilbert space and corresponding
eigenvalues for $x^4$-anharmonic oscillator. The author constructed the
creation and annihilation operators which diagonalize the Hamiltonian and
showed that, unlike the standard harmonic oscillator, these operators obey
the noncanonical commutation relation. This commutation relation will be the
starting point of our analysis to construct the generalized intelligent
states (generalized squeezed and coherent states) for nonlinear oscillator.
Remember that the generalized intelligent states has been discussed in the
literature for the $SU(2)$ and $SU(1,1)$ simple Lie groups $\left[
18-20\right] $ and for the supersymmetric oscillator $\left[ 21\right] .$

The main purpose of this paper is to consider some general properties of
generalized intelligent states corresponding to $x^4$-nonlinear oscillator.
These states generalize those defined by Gazeau and Klauder $\left[
12\right] $ and minimize the Robertson-Schr\"odinger uncertainty relation $%
\left[ 22-24\right] $. The paper is organized as follows: In section 2, the
Gazeau-Klauder coherent states (eigenvectors of the annihilation operator)
are analyzed. Their properties (continuity in the labeling, temporal
stability, overcompleteness and action identity) are discussed. We show also
that there exists a dynamical algebra, generated by the lowering and raising
(creation and annihilation) operators, which arise in the analysis of
spectral structure of the $x^4$-anharmonic oscillator. This dynamical
algebra seems to be a deformation of the Weyl-Heisenberg algebra and is
isomorphic to $su(1,1)$ algebra. The harmonic oscillator limit is discussed
in the end of this section. Section 3 is devoted to the construction of even
and odd Gazeau-Klauder coherent states for the system under consideration.
The real and imaginary Schr\"odinger Cat states are also constructed using
the result of the section 2. In section 4, by means of the
Robertson-Schr\"odinger uncertainty relation, we construct explicitly the
so-called generalized intelligent states. We discuss the analytical
representation and we give an analytical realization of the generalized
intelligent states. Conclusion and perspectives are summarized in the last
section.

\section{Gazeau-Klauder Coherent states for $\mathbf{x}^4$-\protect\\ %
anharmonic oscillator}

\subsection{Structure of $x^4$-anharmonic oscillator eigenvalues}

In this section, we recall the general structure of the energy
eigenvalues for one-dimensional nonlinear oscillators. To be
specific, we are interested in the Hamiltonian which has the form

\begin{equation}
H=a^{+}a^{-}+\frac I2+\frac \varepsilon 4\left( a^{-}+a^{+}\right)
^4-e_0 \label{az},
\end{equation}
where $a^{+}$ and $a^{-}$ are the creation and annihilation operators for
the harmonic oscillator and $I$ is the identity operator. The parameter $%
\varepsilon $ is positive. The quantity $e_0$ is defined as follows

\begin{equation}
e_0=\frac 12+\frac 34\varepsilon -\frac{21}8\varepsilon ^2
\label{sd}.
\end{equation}
The Hamiltonian $H$ can be factorized in the following form $\left[
17\right] $

\begin{equation}  \label{ae}
H=A^{+}A^{-}
\end{equation}
in terms of the new operators $A^{-}$ and $A^{+}$ ($\left( A^{+}\right)
^{\dagger }=A^{-}$) which are defined by

\begin{equation}  \label{de}
A^{-}=a^{-}+\frac \varepsilon 4f_1\left( a^{-},a^{+}\right) +\frac{%
\varepsilon ^2}2f_2\left( a^{-},a^{+}\right),
\end{equation}
where the functions $f_1$ and $f_2$ of the $a^{-}$ and $a^{+}$ are given by

\begin{equation}
f_1\left( a^{-},a^{+}\right) =-3\left( a^{-2}-a^{+2}\right) a^{-}+\left(
a^{-}+a^{+}\right) ^3+3\left( a^{-}+a^{+}\right)  \label{wx}
\end{equation}
and
\begin{eqnarray}
f_2\left( a^{-},a^{+}\right) &=&\frac 32a^{-5}+\frac{39}4a^{+}a^{-4}+\frac{25%
}8a^{+2}a^{-3}-12a^{+3}a^{-2}-\frac 38a^{+4}a^{-}+\frac 14a^{+5}
\\
&&+\frac{75}4a^{-3}-\frac{135}8a^{+}a^{-2}-\frac{135}4a^{+2}a^{-}-\frac
38a^{+3}-\frac{135}8a^{-}-\frac{27}2a^{+}  \nonumber.
\end{eqnarray}
The energy levels are given by $\left[ 17\right] $ (see also $\left[
16\right] $)

\begin{equation}
e_n=n+\frac 32\varepsilon \left( n^2+n\right)  \label{wv}.
\end{equation}
We keep terms only up to $\varepsilon $ which is the standard first-order
perturbation result. The limit $\varepsilon \rightarrow 0$ give the standard
harmonic oscillator. The energy levels constitute a strictly increasing
sequence of positive numbers. The Hilbert space $\mathcal{H}$ for $x^4$%
-non-linear oscillator is easily constructed in the same way as the standard
harmonic oscillator. This space is spanned by the states

\begin{equation}
\left| n,\varepsilon \right\rangle =\frac{\left( A^{+}\right) ^n}{\sqrt{F(n)}%
}e^{i\alpha e_n}\left| 0,\varepsilon \right\rangle,%
\hspace{1cm}n\in \mathbf{N}  \label{wb}
\end{equation}
where $\left| 0,\varepsilon \right\rangle $ is the ground state and the
function $F(n)$ is defined by

\begin{equation}
F(n)=\left\{
\begin{array}{c}
\hspace{0.5cm}1\hbox{\hspace{2.2cm}if }\hspace{1cm}n=0 \\
e_1e_2...e_n\hbox{\hspace{1,5cm}if \hspace{1cm}}n\neq 0
\end{array}
\right.  \label{wn}.
\end{equation}
The action of the annihilation and creation operators are defined as follows:

\begin{eqnarray}
A^{+}\left| n,\varepsilon \right\rangle &=&\sqrt{e_{n+1}}e^{-i\alpha \left(
e_{n+1}-e_n\right) }\left| n+1,\varepsilon \right\rangle  \nonumber \\
A^{-}\left| n,\varepsilon \right\rangle &=&\sqrt{e_n}e^{i\alpha \left(
e_n-e_{n-1}\right) }\left| n-1,\varepsilon \right\rangle  \label{qs}
\end{eqnarray}
where $\alpha \in \mathbf{R}$. We define the number operator $N$ as
\begin{equation}
N\left| n,\varepsilon \right\rangle =n\left| n,\varepsilon
\right\rangle \label{qd}.
\end{equation}
The operator $N$ is different from the product $A^{+}A^{-}(=H)$.

\subsection{Gazeau-Klauder coherent states}

It is well known that, for harmonic oscillator, there are three equivalent
definitions for the coherent states. One consists of defining them as
eigenstates of the annihilation operator of the system. Another possibility
is to define them as the vectors resulting from the application of the
unitary displacement operator on an extremal state (the ground state in
general). A third definition characterizes the coherent states as
minimum-uncertainty states. Recently, Gazeau and Klauder defined the
coherent states, for an arbitrary quantum system, as eigenvectors of the
annihilation operator. Using this definition, the coherent states for $x^4$%
-anharmonic oscillator will be constructed in what follows.

Let us denote the coherent states $\left| z,\alpha \right\rangle $
to show explicitly their dependance on the parameter $\alpha $
(the relevance of this parameter will be clear afterwards (see
equations $\left( 15\right) $
and $\left( 31\right) $). They are eigenstates of $A^{-}$%
\begin{equation}
A^{-}\left| z,\alpha \right\rangle =z\left| z,\alpha \right\rangle
\label{qf}.
\end{equation}
To get their explicit form, we decompose $\left| z,\alpha \right\rangle $ in
terms of the $\left| n,\varepsilon \right\rangle $ basis
\begin{equation}
\left| z,\alpha \right\rangle =\sum\limits_{n=0}^\infty a_n\left|
n,\varepsilon \right\rangle  \label{qg}.
\end{equation}
Substituting this expression in $\left( 12\right) $ and using $\left(
10\right) $, we get the coefficients $a_n$%
\begin{equation}
a_n=\frac{z^n\hbox{ }e^{-i\alpha
e_n}}{\sqrt{F(n)}}a_0,\hspace{1cm}n\in \mathbf{N}  \label{qh}.
\end{equation}
Then, the normalized coherent states are given by
\begin{equation}
\left| z,\alpha \right\rangle =a_0\sum\limits_{n=0}^\infty \sqrt{\frac{%
\Gamma \left( 2+\frac 2{3\varepsilon }\right) }{\left( 3\varepsilon \right)
^n\Gamma \left( n+1\right) \Gamma \left( n+2+\frac 2{3\varepsilon }\right) }}%
\left( z\sqrt{2}\right) ^ne^{-i\alpha e_n}\left| n,\varepsilon \right\rangle
\label{qj}
\end{equation}
where
\begin{equation}
a_0\equiv a_0\left( \left| z\right| \right) =\left[ \hbox{
}_0F_1\left( 2+\frac 2{3\varepsilon },\frac 2{3\varepsilon }\left|
z\right| ^2\right) \right] ^{-\frac 12}  \label{qk}
\end{equation}
and $_0F_1\left( \beta ,x\right) $ is the hypergeometric function defined
by:
\begin{equation}
_0F_1\left( \beta ,x\right) =\sum\limits_{n=0}^\infty \frac{\Gamma
\left( \beta \right) }{\Gamma \left( \beta +n\right)
}\frac{x^n}{n!}  \label{ql}.
\end{equation}
We observe that the coherent states $\left| z,\alpha \right\rangle $ are
continuously labeled by $z$ and $\alpha $, and exist only if the radius of
convergence
\begin{equation}
\mathcal{R}=\lim\limits_{n\rightarrow \infty }\sqrt[n]{F(n)}  \label{qm}
\end{equation}
is non zero. In our case, this radius is infinite.\\Let us analyze now the
completeness of the set $\{\left| z,\alpha \right\rangle ,$ $z\in \mathbf{C}$
and $\alpha \in \mathbf{R}\}$. We impose
\begin{equation}
\int \left| z,\alpha \right\rangle \left\langle z,\alpha \right| \hbox{ }%
d\mu \left( z\right) =I_{\mathcal{H}}  \label{qw},
\end{equation}
where the measure $d\mu \left( z\right) $ has to be determined. If we
suppose that $d\mu \left( z\right) $ depends only on $\left| z\right| $, it
can be determined as in $\left[ 6\right] $. Indeed, let us take
\begin{equation}
d\mu \left( z\right) =\left[ _0F_1\left( 2+\frac 2{3\varepsilon
},\frac 2{3\varepsilon }\left| z\right| ^2\right) \right] h\left(
r^2\right) rdrd\phi ,\hspace{1cm}z=re^{i\phi }  \label{qx}.
\end{equation}
Then, performing the integral in the angular variable $\phi $, we get
\begin{equation}
I=\sum\limits_{n=0}^\infty \left| n,\varepsilon \right\rangle
\left\langle
n,\varepsilon \right| \left[ \frac \pi {F(n)}\int_0^{+\infty }x^n\hbox{ }%
h(x)\hbox{ }dx\right]  \label{qc}.
\end{equation}
In order to recover the resolution of the identity in terms of the basis $%
\left\{ \left| n,\varepsilon \right\rangle \right\} $ we must have
\begin{equation}
\int_0^{+\infty }y^{n-1}\hbox{ }g(y)\hbox{ }dy=\Gamma \left(
n\right) \Gamma \left( n+\frac 2{3\varepsilon }+1\right)
\label{qv},
\end{equation}
where
\begin{equation}
g(y)=\frac{3\pi \varepsilon }2\hbox{ }h\left( \frac{3\varepsilon
}2y\right) \Gamma \left( 2+\frac 2{3\varepsilon }\right)
\label{qb}.
\end{equation}
Then it is clear that the function we are looking for is the
inverse Mellin transform of $g(s)=\Gamma (s)\Gamma \left( s+\frac
2{3\varepsilon
}+1\right) $, $s\in \mathbf{C}$.\\The result is $\left[ 25\right] $%
\begin{equation}
h(x)=\frac{4\left( \sqrt{\frac 2{3\varepsilon }x}\right) ^{\left(
\frac 2{3\varepsilon }+1\right) }}{3\pi \varepsilon \Gamma \left(
2+\frac 2{3\varepsilon }\right) }K_{\left( \frac 12+\frac
1{3\varepsilon }\right) }\left( 2\sqrt{\frac 2{3\varepsilon
}x}\right)  \label{qn},
\end{equation}
where
\begin{equation}
K_\upsilon (z)=\frac \pi 2\frac{I_{-\upsilon }(z)-I_\upsilon (z)}{\sin
(\upsilon z)}  \label{qa}
\end{equation}
is the modified Bessel function of the third kind and
\begin{equation}
I_\upsilon (z)=\sum\limits_{n=0}^\infty \frac{\left( \frac
z2\right) ^{\upsilon +2n}}{n!\hbox{ }\Gamma (n+\upsilon +1)}
\label{qz}
\end{equation}
is the modified Bessel function of the first kind.\\There are two main
consequences arising from the former result. First, we can express any
coherent state in terms of the others
\begin{equation}
\left| z^{\prime },\alpha \right\rangle =\int \left| z,\alpha
\right\rangle \left\langle z,\alpha \right. \left| z^{\prime
},\alpha \right\rangle \hbox{ }d\mu (z)  \label{qe}.
\end{equation}
The kernel $\left\langle z,\alpha \right. \left| z^{\prime },\alpha
\right\rangle $ is
\begin{equation}
\left\langle z,\alpha \right. \left| z^{\prime },\alpha \right\rangle =\frac{%
_0F_1\left( 2+\frac 2{3\varepsilon },\frac 2{3\varepsilon }\overline{z}%
z^{\prime }\right) }{\sqrt{_0F_1\left( \frac 2{3\varepsilon
}+2,\frac 2{3\varepsilon }\left| z\right| ^2\right) _0F_1\left(
\frac 2{3\varepsilon }+2,\frac 2{3\varepsilon }\left| z^{\prime
}\right| ^2\right) }}  \label{qr}.
\end{equation}
Second, an arbitrary element $\left| g,\varepsilon \right\rangle
=\sum\limits_{m=0}^\infty b_m\left| m,\varepsilon \right\rangle $, of the%
\textit{\ }Hilbert space $\mathcal{H}$, can be written in terms of the
coherent states as
\begin{equation}
\left| g,\varepsilon \right\rangle =\int \left\langle z,\alpha
\right.
\left| g,\varepsilon \right\rangle \left| z,\alpha \right\rangle \hbox{ }%
d\mu (z)  \label{qt},
\end{equation}
where
\begin{equation}
\left\langle z,\alpha \right. \left| g,\varepsilon \right\rangle
=a_0\sum\limits_{n=0}^\infty \sqrt{\frac{\Gamma \left( 2+\frac
2{3\varepsilon }\right) }{\left( 3\varepsilon \right) ^n\Gamma
\left( n+1\right) \Gamma \left( n+2+\frac 2{3\varepsilon }\right)
}}\left( \overline{z}\sqrt{2}\right) ^nb_ne^{i\alpha e_n}
\label{qy}
\end{equation}
determine completely the state $\left| g,\varepsilon \right\rangle \in
\mathcal{H}$

Let us now consider the dynamical evolution of the coherent states, which is
quite simple. Indeed, we have
\begin{equation}
U\left( t\right) \left| z,\alpha \right\rangle =e^{-itH}\left|
z,\alpha \right\rangle =\left| z,\alpha +t\right\rangle
\label{qu}.
\end{equation}
It is clear from equations $\left( 12\right) $, $\left( 15\right) ,$ $\left(
19\right) $ and $\left( 31\right) $ that the $x^4$-anharmonic oscillator
coherent states constructed as eigenvectors of the annihilation operator $%
A^{-}$ satisfy the following properties

\hspace{1cm}(i) Continuity : $z\in \mathbf{C}$ , $\alpha \in \mathbf{R}%
\rightarrow \left| z,\alpha \right\rangle $ is continuous

\hspace{1cm}(ii) The resolution of unity : $I_{\mathcal{H}}=\int $
$\left| z,\alpha \right\rangle \left\langle z,\alpha \right| $
$d\mu (z)$

\hspace{1cm}(iii) Temporal stability : $e^{-itH}\left| z,\alpha
\right\rangle =\left| z,\alpha +t\right\rangle $

\hspace{1cm}(iv) Action identity : $\left\langle z,\alpha \right| H\left|
z,\alpha \right\rangle =\left| z\right| ^2$\\These properties (i) - (iv) are
the set of requirements to define coherent states of an arbitrary quantum
system imposed by Gazeau and Klauder $\left[ 12\right] .$

\subsection{The dynamical algebra : Extended Weyl-Heisenberg\protect\\algebra
}

Behind the spectral structure of the $x^4$-anharmonic oscillator, there
exists a dynamical algebra generated by the lowering and raising operators $%
A^{-}$ and $A^{+}$. We follow a similar approach of the references
$\left[ 14\right] $ and $\left[ 15\right] $in which the authors
discussed the dynamical algebra $su(1,1)$ for the infinite square
well and P\"oschl-Teller Hamiltonians.\\For the $x^4 $-anharmonic
oscillator, we defined the operator number $N$ by
\begin{equation}
N\left| n,\varepsilon \right\rangle =n\left| n,\varepsilon \right\rangle
\label{qi}
\end{equation}
which can be given, formally, in terms of the Hamiltonian $H$ by
\begin{equation}
N=\sqrt{\frac 2{3\varepsilon }H+\left( \frac 12+\frac
1{3\varepsilon }\right) ^2}-\left( \frac 12+\frac 1{3\varepsilon
}\right)  \label{qo}.
\end{equation}
The creation and annihilation operators $A^{+}$ and $A^{-}$ satisfy the
following commutation relations

\begin{equation}
\left[ A^{-},A^{+}\right] =1+3\varepsilon \hbox{ }\left( N+1\right), %
\hspace{1,5cm}\left[ N,A^{\pm }\right] =\pm A^{\pm } \label{qp}.
\end{equation}
In the limit $\varepsilon \rightarrow 0,$ we get the well known
Weyl-Heisenberg algebra. The operators $A^{-}$, $A^{+}$ and $N$ generate an
extended Weyl Heisenberg algebra.\\We note that there exists a dynamical Lie
algebra, which is generated by the new set of operators $\left\{ \tilde
A^{-},\tilde A^{+},\tilde N\right\} $ defined as
\begin{equation}
\tilde A^{\pm }=\sqrt{\frac 2{3\varepsilon }}A^{\pm
},\hspace{1.5cm}\tilde N=2\left( N+1\right) +\frac 2{3\varepsilon
}  \label{qs}
\end{equation}
and satisfying the following commutation relations
\begin{equation}
\left[ \tilde A^{-},\tilde A^{+}\right] =\tilde
N,\hspace{1.5cm}\left[ \tilde A^{\pm },\tilde N\right] =\mp
2\tilde A^{\pm }  \label{qd}
\end{equation}
which is isomorphic to $su(1,1)\sim sl(2,R)\sim so(2,1).$ A more
familiar basis for $su(1,1)$ is given by
\begin{equation}
J_{-}=\frac 1{\sqrt{2}}\tilde A^{-},\hspace{1cm}J_{+}=\frac
1{\sqrt{2}}\tilde A^{+},\hspace{1cm}\hbox{ }J_{12}=\frac 12\tilde
N \label{qf},
\end{equation}
where $J_{12}$ can be seen as the generator of the compact subgroup $SO(2)$,
i.e.,
\begin{equation}
\left[ J_{-},J_{+}\right] =J_{12},\hspace{1.5cm}\left[ J_{\pm
},J_{12}\right] =\mp J_{\pm }  \label{qg}.
\end{equation}
It follows from the above considerations that the space $\mathcal{H}$ of
states $\left| n,\varepsilon \right\rangle $ carries some representation of $%
su(1,1)$. The actions of the Lie algebra elements read
\begin{eqnarray}
J_{+}\left| n,\varepsilon \right\rangle &=&\frac
1{\sqrt{2}}\sqrt{\left( n+1\right) \left( \frac 2{3\varepsilon
}+n+2\right) }e^{i\alpha \left( 3\varepsilon n+3\varepsilon
+1\right) }\left| n+1,\varepsilon \right\rangle \label{qj} ,\\
J_{-}\left| n,\varepsilon \right\rangle &=&\frac
1{\sqrt{2}}\sqrt{\left( n\right) \left( \frac 2{3\varepsilon
}+n+1\right) }e^{-i\alpha \left( 3\varepsilon n+1\right) }\left|
n-1,\varepsilon \right\rangle  \label{qh} ,\\ J_{12}\left|
n,\varepsilon \right\rangle &=&\left( \frac 1{3\varepsilon
}+n+1\right) \left| n,\varepsilon \right\rangle  \label{qk}.
\end{eqnarray}
The $su(1,1)$ casimir operator is
\begin{equation}
C=2J_{+}J_{-}-J_{12}\left( J_{12}-1\right)  \label{ql}
\end{equation}
with the following eigenvalue
\begin{equation}
C\left| n,\varepsilon \right\rangle =\frac 1{3\varepsilon }\left(
\frac 1{3\varepsilon }+1\right) \left| n,\varepsilon \right\rangle
\label{qm}.
\end{equation}
Finally, we note that $su(1,1)$ is the\textit{\ only }dynamical Lie algebra
that can arise in such a problem.

It is clear that the $x^4$- anharmonic oscillator tends to the harmonic
oscillator Hamiltonian in the limit $\varepsilon \rightarrow 0$. Let us
consider this limit in detail to see that the coherent states we have
computed give the harmonic oscillator ones.\\Because
\begin{equation}
\lim\limits_{\varepsilon \rightarrow 0}\frac{2^n\Gamma \left(
2+\frac 2{3\varepsilon }\right) }{\left( 3\varepsilon \right)
^n\Gamma \left( n+2+\frac 2{3\varepsilon }\right) }=1  \label{sa},
\end{equation}
one can see that
\begin{equation}
\lim\limits_{\varepsilon \rightarrow 0}\hbox{ }\left[ a_0\left(
\left| z\right| \right) \right] ^{-2}=\sum\limits_{n=0}^\infty
\frac{\left| z\right| ^{2n}}{\Gamma (n+1)}=e^{\left| z\right| ^2}
\label{sz}
\end{equation}
and
\begin{equation}
\lim\limits_{\varepsilon \rightarrow 0}\left| z,\alpha \right\rangle =e^{-%
\frac{\left| z\right| ^2}2}\sum\limits_{n=0}^\infty \frac{z^n}{\sqrt{n!}}%
\hbox{ }e^{-in\alpha }\left| n,0\right\rangle  \label{se},
\end{equation}
where $\left| n,0\right\rangle $ is the number state for the usual
harmonic oscillator ($\left| n,0\right\rangle \equiv \left|
n,\varepsilon \right\rangle $ for $\varepsilon \rightarrow 0$).

\section{Superpositions of Gazeau-Klauder Coherent states}

\subsection{Even and odd coherent states}

Even and odd coherent states give rise to states with non classical
properties. An important case is the superposition of the coherent states $%
\left| z,\alpha \right\rangle $ and $\left| -z,\alpha \right\rangle $ and
the resultant states are eigenvalues of the operator $\left( A^{-}\right) ^2$%
. Here, we show that this class of even and odd coherent states can be
generated for the $x^4$- anharmonic oscillator.

Using the Gazeau-Klauder coherent states $\left( 15\right) $ for $x^4$%
-anharmonic oscillator, we define the even coherent states by the symmetric
combination
\begin{equation}
\left| z,\alpha \right\rangle _e=\frac{\mathcal{N}_e}{2a_0}\hbox{
}\left( \left| z,\alpha \right\rangle +\left| -z,\alpha
\right\rangle \right)
\end{equation}
and odd coherent states by the antisymmetric combination
\begin{equation}
\left| z,\alpha \right\rangle _o=\frac{\mathcal{N}_o}{2a_0}\hbox{
}\left( \left| z,\alpha \right\rangle -\left| -z,\alpha
\right\rangle \right),
\end{equation}
where the constant $a_0$ is given by equation $\left( 16\right) .$\\It is
easy to derive the corresponding expansions for even and odd coherent states

\begin{equation}
\left| z,\alpha \right\rangle
_e=\mathcal{N}_e\sum\limits_{k=0}^\infty \sqrt{\frac{\Gamma \left(
2+\frac 2{3\varepsilon }\right) }{\left( 3\varepsilon \right)
^{2k}\Gamma \left( 2k+1\right) \Gamma \left( 2k+2+\frac
2{3\varepsilon }\right) }}\left( z\sqrt{2}\right) ^{2k}e^{-i\alpha
e_{2k}}\left| 2k,\varepsilon \right\rangle,
\end{equation}
and
\begin{equation}
\left| z,\alpha \right\rangle
_o=\mathcal{N}_o\sum\limits_{k=0}^\infty \sqrt{\frac{\Gamma \left(
2+\frac 2{3\varepsilon }\right) }{\left( 3\varepsilon \right)
^{2k+1}\Gamma \left( 2k+2\right) \Gamma \left( 2k+3+\frac
2{3\varepsilon }\right) }}\left( z\sqrt{2}\right)
^{2k+1}e^{-i\alpha e_{2k+1}}\left| 2k+1,\varepsilon \right\rangle.
\end{equation}
The normalization constant $\mathcal{N}_e$ and $\mathcal{N}_o$ can be
obtained from the normalization conditions
\begin{equation}
_{\stackrel{}{e}}\left\langle z,\alpha \right. \left| z,\alpha
\right\rangle _e=1,\hspace{2cm}_o\left\langle z,\alpha \right.
\left| z,\alpha \right\rangle _o=1
\end{equation}
which leads to
\begin{equation}
\mathcal{N}_e=\mathcal{N}_e\mathcal{(}\left| z\right| )=\left[ \hbox{ }%
_0F_3\left( \frac 12,\frac 1{3\varepsilon }+1,\frac 1{3\varepsilon }+\frac
32,\frac{\left| z\right| ^4}{\left( 6\varepsilon \right) ^2}\right) \right]
^{-\frac 12}
\end{equation}
and
\begin{equation}
\mathcal{N}_o=\mathcal{N}_o\mathcal{(}\left| z\right| )=\left[
\frac{\left| z\right| ^2}{\left( 1+3\varepsilon \right) }\hbox{
}_0F_3\left( \frac 32,\frac 1{3\varepsilon }+2,\frac
1{3\varepsilon }+\frac 32,\frac{\left| z\right| ^4}{\left(
6\varepsilon \right) ^2}\right) \right] ^{-\frac 12}.
\end{equation}
It is straighfoward to show that the even (or odd) coherent states cannot
form separately a complete set. However the even coherent states together
with the odd ones build an overcomplete Hilbert space. Their completeness
relation takes the form
\begin{equation}
\int \left| z,\alpha \right\rangle _{o}\hbox{ }_{\stackrel{}{%
o}}\left\langle z,\alpha \right| \hbox{ }d\mu _o(z)+\int \left|
z,\alpha \right\rangle _{e}\hbox{ }_{\stackrel{}{e}}\left\langle
z,\alpha \right| \hbox{ }d\mu _e(z)=I_{\mathcal{H}},
\end{equation}
where the two weight function are given by

\begin{equation}
d\mu _o(z)=\left[ \mathcal{N}_o\right] ^{-2}h\left( r^2\right) rdrd\phi, %
\hspace{1cm}z=re^{i\phi }
\end{equation}
and
\begin{equation}
d\mu _e(z)=\left[ \mathcal{N}_e\right] ^{-2}h\left( r^2\right) rdrd\phi, %
\hspace{1cm}z=re^{i\phi }
\end{equation}
where $h\left( r^2\right) $ in given by equation $\left( 24\right) .$\\The
even and odd coherent states have the following orthogonal relations
\begin{eqnarray}
_{\stackrel{}{e}}\left\langle z,\alpha \right. \left| z^{\prime
},\alpha \right\rangle _e &=&\frac{_0F_3\left( \frac 12,\frac
1{3\varepsilon }+1,\frac 1{3\varepsilon }+\frac 32,\frac{\left(
\overline{z}z^{\prime }\right) ^2}{\left( 6\varepsilon \right)
^2}\right) }{_0F_3\left( \frac 12,\frac 1{3\varepsilon }+1,\frac
1{3\varepsilon }+\frac 32,\frac{\left| z\right| ^4}{\left(
6\varepsilon \right) ^2}\right) } ,\\
_{\stackrel{}{o}}\left\langle z,\alpha \right. \left| z^{\prime
},\alpha \right\rangle _o &=&\frac{\overline{z}z^{\prime }\hbox{
}_0F_3\left( \frac 32,\frac 1{3\varepsilon }+2,\frac
1{3\varepsilon }+\frac 32,\frac{\left(
\overline{z}z^{\prime }\right) ^2}{\left( 6\varepsilon \right) ^2}\right) }{%
\left| z\right| ^2\hbox{ }_0F_3\left( \frac 32,\frac
1{3\varepsilon }+2,\frac 1{3\varepsilon }+\frac 32,\frac{\left|
z\right| ^4}{\left( 6\varepsilon \right) ^2}\right) } ,\\
_{\stackrel{}{o}}\left\langle z,\alpha \right. \left| z^{\prime
},\alpha \right\rangle _e &=&0
\end{eqnarray}
where $\overline{z}$ is the complex conjugation of $z$.

\subsection{Probability distribution of even and odd Gazeau-Klauder coherent
states}

The probability distribution of even and odd Gazeau-Klauder coherent states
for $x^4$-anharmonic oscillator is defined as follow

\begin{equation}
\mathcal{P}_e\left( n,\varepsilon \right) =\left| \left\langle n,\varepsilon
\right. \left| z,\alpha \right\rangle _e\right| ^2,\hspace{1.0in}\mathcal{P}%
_o\left( n,\varepsilon \right) =\left| \left\langle n,\varepsilon
\right. \left| z,\alpha \right\rangle _o\right| ^2.
\end{equation}
Using equations $\left( 49\right) $ and $\left( 50\right) $ we
obtain

\begin{equation}
\mathcal{P}_e\left( n,\varepsilon \right) =\mathcal{P}_e\left(
2k,\varepsilon \right) =\frac{\mathcal{N}_e^2\left| z\right|
^{4k}}{F(2k)},
\end{equation}
and
\begin{equation}
\mathcal{P}_o\left( n,\varepsilon \right) =\mathcal{P}_o\left(
2k+1,\varepsilon \right) =\frac{\mathcal{N}_o^2\left| z\right| ^{4k+2}}{%
F(2k+1)}.
\end{equation}
In the $\varepsilon \rightarrow 0$ limit, one obtain the probability
distribution of even and odd harmonic oscillator states. In equations $%
\left( 61\right) $ and $\left( 62\right) $, the function $F$ is defined by $%
\left( 9\right) $.

\subsection{Real and imaginary Schr\"odinger Cat states}

The real and imaginary Schr\"odinger Cat states constitute another
representative of superposition states
\begin{equation}
\left| z,\alpha \right\rangle _{+}=\frac{\mathcal{N}_{+}}{2a_0}\hbox{ }%
\left( \left| z,\alpha \right\rangle +\left| \overline{z},\alpha
\right\rangle \right),
\end{equation}
and
\[
\left| z,\alpha \right\rangle _{-}=\frac{\mathcal{N}_{-}}{2ia_0}\hbox{ }%
\left( \left| z,\alpha \right\rangle -\left| \overline{z},\alpha
\right\rangle \right),\]
where $\left| z,\alpha \right\rangle $ is
the coherent states given by equation $\left( 15\right) $ and
$\mathcal{N}_{\pm }$ are the normalization constants of real and
imaginary Schr\"odinger Cat states. Similar superpositions were
considered by Dodonov and al $\left[ 26\right] $ for the usual
harmonic oscillator coherent states, and Roy $\left[ 27\right] $
for the nonlinear coherent states of the centre of mass of a
trapped and biochromatically laser driven far from the Lamb-Dicke
regime.

A direct computation leads to the following expansions for the real and
imaginary Cat states
\begin{equation}
\left| z,\alpha \right\rangle
_{+}=\mathcal{N}_{+}\sum\limits_{n=0}^\infty \frac{r^n\cos \left(
n\phi \right) }{\sqrt{F(n)}}e^{-i\alpha e_n}\left| n,\varepsilon
\right\rangle ,\hspace{1cm}z=re^{i\phi }
\end{equation}
and
\begin{equation}
\left| z,\alpha \right\rangle
_{-}=\mathcal{N}_{-}\sum\limits_{n=0}^\infty
\frac{r^{n+1}\sin \left( \left( n+1\right) \phi \right) }{\sqrt{F(n+1)}}%
e^{-i\alpha e_{n+1}}\left| n+1,\varepsilon \right\rangle.
\end{equation}
The factors $\mathcal{N}_{+}$and $\mathcal{N}_{-}$ given by
\begin{equation}
\mathcal{N}_{+}=\mathcal{N}_{+}\left( \left| z\right| \right) =\left(
\sum\limits_{n=0}^\infty \frac{r^{2n}\cos ^2\left( n\phi \right) }{F(n)}%
\right) ^{-\frac 12}
\end{equation}
and
\begin{equation}
\mathcal{N}_{-}=\mathcal{N}_{-}\left( \left| z\right| \right)
=\left( \sum\limits_{n=0}^\infty \frac{r^{2n+2}\sin ^2\left(
\left( n+1\right) \phi \right) }{F(n+1)}\right) ^{-\frac 12}
\end{equation}
are the normalization constants.

\subsection{Probability distribution of real and Imaginary nonlinear Schr\"odinger Cat states}

The probability distribution of Cat states associated with $x^4$-anharmonic
oscillator is defined as follow
\begin{equation}
\mathcal{P}_{\pm }\left( n\right) =\left| \left\langle
n,\varepsilon \right. \left| z,\alpha \right\rangle _{\pm }\right|
^2.
\end{equation}
Using equations $\left( 64\right) $ and $\left( 65\right) $, we
obtain
\begin{equation}
\mathcal{P}_{\pm }\left( n\right) =\frac{r^{2n}\left( 1\pm \cos
\left( 2n\phi \right) \right) }{A_{\pm }},
\end{equation}
where
\begin{eqnarray}
A_{+} &=&F(n)\left[ \sum\limits_{m=0}^\infty \frac{r^{2m}\left(
1+\cos \left( 2m\phi \right) \right) }{F(m)}\right]  \nonumber \\
A_{-} &=&F(n)\left[ \sum\limits_{m=0}^\infty \frac{r^{2m+2}\left(
1+\cos \left( \left( 2m+2\right) \phi \right) \right)
}{F(m+1)}\right].
\end{eqnarray}
It is clear that, in the limit $\varepsilon \rightarrow 0,$ the equation $%
\left( 69\right) $ reproduce the probability distribution of Cat states for
the harmonic oscillator.

\section{Robertson-Schr\"odinger uncertainty relation}

Using the creation and annihilation operators $A^{+}$ and $A^{-}$, we
introduce two hermitian operators $X$ and $P$%
\begin{equation}
X=\frac 1{\sqrt{2}}\left( A^{-}+A^{+}\right) ,\hspace{2cm}P=\frac i{\sqrt{2}%
}\left( A^{+}-A^{-}\right),
\end{equation}
which satisfy the following commutation relation
\begin{equation}
\left[ X,P\right] =iG\left( N\right) \equiv iG,
\end{equation}
where the operator $G\left( N\right) $ is given by
\begin{equation}
G\left( N\right) =1+3\varepsilon (N+1).
\end{equation}
For $\varepsilon \neq 0$, the operator $G$ is not a multiple of the unit
operator. Then, for the operators $X$ and $P$ satisfying the commutation
relation $\left( 72\right) $, the variances $\left( \Delta X\right) ^2$ and $%
\left( \Delta P\right) ^2$ obey to the Robertson-Schr\"odinger uncertainty
relation
\begin{equation}
\left( \Delta X\right) ^2\left( \Delta P\right) ^2\geq \frac
14\left( \left\langle G\right\rangle ^2+\left\langle
C\right\rangle ^2\right),
\end{equation}
where the operator $C$ is defined by
\begin{equation}
C=\left\{ X-\left\langle X\right\rangle ,P-\left\langle P\right\rangle
\right\}
\end{equation}
or, in terms of the operators $A^{-}$ and $A^{+}$, by
\begin{equation}
C=i\left[ \left( 2A^{-}-\left\langle A^{-}\right\rangle \right)
\left\langle A^{-}\right\rangle +\left( -2A^{+}+\left\langle
A^{+}\right\rangle \right) \left\langle A^{+}\right\rangle
-A^{-2}+A^{+2}\right].
\end{equation}
The symbol $\left\{ \hbox{ },\hbox{ }\right\} $ stands for the
anticommutator. When there is a correlation between $X$ and $P$ i.e., $%
\left\langle C\right\rangle \neq 0,$ such a relation is a generalization of
the usual one (Heisenberg uncertainty relation)
\begin{equation}
\left( \Delta X\right) ^2\left( \Delta P\right) ^2\geq \frac
14\left\langle G\right\rangle ^2.
\end{equation}
The special form $\left( 77\right) $ is, of course, identical with the
general form $\left( 74\right) $ if $X$ and $P$ are uncorrelated, i.e. $%
\left\langle C\right\rangle =0$\\The general uncertainty relation $\left(
74\right) $ is better suited to determine the lower bound on the product of
variances in the measurement of observables corresponding to noncanonical
operators. The Robertson-Schr\"odinger uncertainty relation give us a new
understanding of what the states are coherent and squeezed for an arbitrary
quantum system $\left[ 28\right] $. Indeed the so called generalized
intelligent states are obtained when the equality in the
Robertson-Schr\"odinger uncertainty relation is realized (see $\left[ 22,%
\hbox{ }28\right] $). The inequality in equation $\left( 74\right)
$ become the equality for the states satisfying the following
eigenvalues equation
\begin{equation}
\left( X+i\lambda P\right) \left| \psi \right\rangle
=z\sqrt{2}\left| \psi \right\rangle, \hspace{1.5cm}\lambda \hbox{
},\hbox{ }z\in \mathbf{C}.
\end{equation}
As a consequence, we have the following relations
\begin{equation}
\left( \Delta X\right) ^2=\left| \lambda \right| \Delta,
\hspace{1.0in}\left( \Delta P\right) ^2=\frac 1{\left| \lambda
\right| }\Delta  \label{zg},
\end{equation}
with
\begin{equation}
\Delta =\frac 12\sqrt{\left\langle G\right\rangle ^2+\left\langle
C\right\rangle ^2}  \label{zh}.
\end{equation}
Note that the average values $\left\langle G\right\rangle $ and $%
\left\langle C\right\rangle $ can be expressed in terms of the variances as
\begin{eqnarray}
\left\langle G\right\rangle &=&2\hbox{Re}\left( \lambda \right)
\left( \Delta P\right) ^2 , \nonumber \\ \left\langle
C\right\rangle &=&2\hbox{Im}\left( \lambda \right) \left( \Delta
P\right) ^2  \label{zh}.
\end{eqnarray}
It is clear, from $\left( 79\right) ,$ that in the case where
$\left| \lambda \right| =1$,

\begin{equation}  \label{zj}
\left( \Delta X\right) ^2=\left( \Delta P\right) ^2,
\end{equation}
and we call the states satisfying $\left( 82\right) $, with $\left| \lambda
\right| =1,$ the generalized coherent states and for $\left| \lambda \right|
\neq 1$, the states are called generalized squeezed states.

Using the equation $\left( 78\right) ,$ one can obtain some general
relations for the average values and dispersions for the operators $X$ and $%
P $ in the states minimizing the Robertson-Schr\"odinger uncertainty
relation $\left( 74\right) $. Indeed, we have $\left[ 18,21\right] $%
\begin{eqnarray}
\left( \Delta X\right) ^2 &=&\frac 12\left( \hbox{Re}\left(
\lambda \right) \left\langle G\right\rangle +\hbox{Im}\left(
\lambda \right) \left\langle C\right\rangle \right)  \label{zk},
\\
\left( \Delta P\right) ^2 &=&\frac 1{2\left| \lambda \right| ^2}\left( \hbox{%
Re}\left( \lambda \right) \left\langle G\right\rangle
+\hbox{Im}\left( \lambda \right) \left\langle C\right\rangle
\right)  \label{zl}, \\
\hbox{Im}\left( \lambda \right) \left\langle G\right\rangle &=&\hbox{Re}%
\left( \lambda \right) \left\langle C\right\rangle  \label{zm}.
\end{eqnarray}
The minimization of the Robertson-Schr\"odinger uncertainty relation leads
to generalized coherent states for $\left| \lambda \right| =1$ (including
Gazeau-Klauder coherent states) and generalized squeezed states for $\left|
\lambda \right| \neq 1.$

\subsection{Generalized Intelligent States for $\mathbf{x}^4$-anharmonic
\protect\\oscillator}

In order to give a complete classification of the coherent and squeezed
states for $x^4$-anharmonic oscillator, we will, in what follows, solve the
eigenvalues equation $\left( 78\right) $. So, using the definition of $X$
and $P$ in terms of the annihilation and creation operators $A^{-}$ and $%
A^{+}$, the eigenvalues equation $\left( 78\right) $ takes the following
form:
\begin{equation}
\left( \left( 1-\lambda \right) A^{+}+\left( 1+\lambda \right)
A^{-}\right) \left| \psi \right\rangle =2z\left| \psi
\right\rangle.
\end{equation}
Let us compute $\left| \psi \right\rangle $ explicitly using $\left(
86\right) $. we take
\begin{equation}
\left| \psi \right\rangle =\sum\limits_{n=0}^\infty c_n\left|
n,\varepsilon \right\rangle.
\end{equation}
So that
\[
\left( 1-\lambda \right) c_{n-1}\sqrt{e_n}e^{-i\alpha \left(
e_n-e_{n-1}\right) }+\left( 1+\lambda \right) c_{n+1}\sqrt{e_{n+1}}%
e^{i\alpha \left( e_{n+1}-e_n\right) }=2zc_n,
\]

\begin{equation}
\left( 1+\lambda \right) \sqrt{e_1}c_1=2ze^{-i\alpha e_1}c_0
\label{ed}.
\end{equation}
The recurrence formulae $\left( 88\right) $ will give a complete
classification of generalized intelligent states corresponding to the system
under consideration.

By taking specific values of $\lambda $ and $z$, we now analyze the solution
arising from the recurrence relation $\left( 88\right) $. We will study the
cases ($\lambda \neq -1$, $z\neq 0$), ($\lambda =1$, $z\neq 0$), ($\lambda
=-1$, $z\neq 0$), and ($\lambda \neq -1$, $z=0$).

We start by examining the general case ($\lambda \neq -1$, $z\neq 0$). To
solve the eigenvalues equation $\left( 86\right) $, we set
\begin{equation}
A_{n+1}=\frac{c_{n+1}}{c_n}\sqrt{e_{n+1}}e^{i\alpha
(e_{n+1}-e_n)}.
\end{equation}
The relation $\left( 88\right) $ can be written in terms of the new
coefficients $A_n$ as follows

\begin{equation}
A_1=\frac{2z}{\left( 1+\lambda \right) },\hspace{0.5cm}%
\hspace{1cm}A_n=\frac{2z}{\left( 1+\lambda \right) }+\left( \frac{\lambda -1%
}{\lambda +1}\right) \frac{e_{n-1}}{A_{n-1}}  \label{qp}.
\end{equation}
By some elementary manipulation, we obtain the coefficients $A_n$ as a
continued fraction given by:

\begin{equation}
A_n=\frac{2z}{1+\lambda }+\frac{\left( \frac{\lambda -1}{\lambda +1}%
\right) e_{n-1}}{\frac{2z}{1+\lambda }+\frac{\left( \frac{\lambda -1}{%
\lambda +1}\right) e_{n-2}}{\frac{2z}{1+\lambda }+\frac{\left( \frac{%
\lambda -1}{\lambda +1}\right) e_{n-3}}{%
\begin{array}{c}
\frac{2z}{1+\lambda }+.....\hspace{4cm} \\ \hspace{0.3cm}.+.....
\\ \hspace{5cm}\frac{2z}{1+\lambda }+\frac{\left( \frac{\lambda
-1}{\lambda +1}\right) e_1}{A_1}.
\end{array}
}}}\hspace{1cm}  \label{wq}
\end{equation}
Using the result $\left( 91\right) $ and the equation $\left( 89\right) $,
we prove that the coefficients $c_n$ are given by

\begin{equation}
c_n=c_0\frac{\left( 2z\right) ^n}{\left( 1+\lambda \right)
^n\sqrt{F\left( n\right) }}\left[ \sum\limits_{h=0\left( 1\right)
\left[ \frac n2\right] }\left( -1\right) ^h\frac{\left( 1-\lambda
^2\right) ^h}{\left( 2z\right) ^{2h}}\Delta \left( n,h\right)
\right] e^{-i\alpha e_n}  \label{cq},
\end{equation}
where the symbol $\left[ \frac n2\right] $ stands for the integer part of $%
\frac n2$ and the function $\Delta \left( n,h\right) $ is defined by

\begin{equation}  \label{vq}
\Delta \left( n,h\right) =\sum\limits_{j_1=1}^{n-\left(
2h-1\right) }e_{j_1}\left[ \sum\limits_{j_2=j_1+2}^{n-\left(
2h-3\right) }e_{j_2}...\left[ ...\left[
\sum\limits_{j_h=j_{h-1}+2}^{n-1}e_{j_h}\right] \right]
...\right].
\end{equation}
The states $\left| \psi \right\rangle =\sum\limits_{n=0}^\infty
c_n\left| n,\varepsilon \right\rangle $ obtained here can be also
written as some operator acting on the ground state $\left|
0,\varepsilon \right\rangle $. Indeed, we have

\begin{equation}
\left| \psi \right\rangle =U(\lambda \neq -1\hbox{, }z\neq
0)\left| 0,\varepsilon \right\rangle,
\end{equation}
where the operator $U(\lambda \neq -1$, $z\neq 0)$ is given by

\begin{equation}
U(\lambda \neq -1,z\neq 0)=c_0\sum\limits_{n=0}^\infty \left( \left( \frac{%
2z}{1+\lambda }\right) \frac 1HA^{+}+\left( \frac{\lambda -1}{\lambda +1}%
\right) \frac 1H\left( A^{+}\right) ^2\right) ^n.
\end{equation}
Taking $\lambda =1$ one can recover the Gazeau-Klauder coherent states (up
to normalization constant)
\begin{equation}
\left| \psi \right\rangle =\exp \left( z\frac NHA^{+}\right) \left|
0,\varepsilon \right\rangle
\end{equation}
given here as the action of the operator $\exp \left( z\frac NHA^{+}\right) $
on the ground state $\left| 0,\varepsilon \right\rangle $. In the limit $%
\varepsilon \rightarrow 0$, we get the generalized intelligent states (up to
normalization constant)
\begin{equation}
\left| \psi \right\rangle =\exp \left( \frac{2z}{1+\lambda }a^{+}\right)
\exp \left( \left( \frac{\lambda -1}{\lambda +1}\right) \frac{\left(
a^{+}\right) ^2}2\right) \left| 0\right\rangle
\end{equation}
corresponding to harmonic oscillator.

It is clear that the equation $\left( 86\right) $ give also the
Gazeau-Klauder\ coherent states. In this case, the variances $\left( \Delta
X\right) ^2$ and $\left( \Delta P\right) ^2$ are given by
\begin{equation}
\left( \Delta X\right) ^2=\left( \Delta P\right) ^2=\frac
12\left\langle G\right\rangle,
\end{equation}
where
\begin{equation}
\left\langle G\right\rangle =\left( 1+3\varepsilon \right) +\frac{%
3\varepsilon \left| z\right| ^2}{1+3\varepsilon }\frac{_0F_1\left( \frac
2{3\varepsilon }+3,\frac 2{3\varepsilon }\left| z\right| ^2\right) }{%
_0F_1\left( \frac 2{3\varepsilon }+2,\frac 2{3\varepsilon }\left|
z\right| ^2\right) }.
\end{equation}
Note that (see the equation $\left( 85\right) $) $\left\langle
C\right\rangle =0$ when $\lambda =1$. Then, the Gazeau-Klauder coherent
states are minimum Heisenberg uncertainty states and verify
\begin{equation}
\left( \Delta X\right) \left( \Delta P\right) =\frac 12\left[ \left(
1+3\varepsilon \right) +\frac{3\varepsilon \left| z\right| ^2}{%
1+3\varepsilon }\frac{_0F_1\left( \frac 2{3\varepsilon }+3,\frac
2{3\varepsilon }\left| z\right| ^2\right) }{_0F_1\left( \frac
2{3\varepsilon }+2,\frac 2{3\varepsilon }\left| z\right| ^2\right)
}\right].
\end{equation}
By taking the limit $\varepsilon \rightarrow 0$, one can see that the
previous formula takes the simpler form
\begin{equation}
\left( \Delta X\right) \left( \Delta P\right) =\frac 12
\end{equation}
which is a standard result.\\Thus, we have found some additional information
concerning the Gazeau-Klauder coherent states. The result obtained here is
true for any (exact solvable model) arbitrary quantum system; see the
reference $\left[ 28\right] $. Indeed, through the analysis of the states
minimizing the Robertson-Schr\"odinger uncertainty relation, we have been
able to show that the Gazeau-Klauder coherent states (the eigenvectors of
the annihilation operator) are the states minimizing the Heisenberg
uncertainty relation.

For ($\lambda =-1,z\neq 0$), we have to solve the eigenvalues equation
\begin{equation}
A^{+}\left| \psi \right\rangle =z\left| \psi \right\rangle.
\end{equation}
Then, the recurrence relation $\left( 88\right) $ rewrite as
\begin{equation}
zc_n=c_{n-1}\sqrt{e_n}e^{-i\alpha \left( e_n-e_{n-1}\right) },\hspace{1cm}%
\hspace{0.5cm}c_0=0.
\end{equation}
The coefficients $c_n$ vanishe and we exclude out this case of our analysis.

In the case where ($\lambda \neq -1$, $z=0$) putting $z=0$ in the equation $%
\left( 95\right) $, the solution is given by the action of the operator

\begin{equation}
U(\lambda \neq -1\hbox{, }z=0)=\exp \left( \frac 12\left( \frac{\lambda -1}{%
\lambda +1}\right) \frac NH\hbox{ }\left( A^{+}\right) ^2\right),
\end{equation}
on the ground state $\left| 0,\varepsilon \right\rangle $ where $H$ is the
Hamiltonian of our system. We rewrite the state $\left| \psi \right\rangle $%
, in a compact form, as follows
\begin{equation}
\left| \psi \right\rangle =U(\lambda \neq -1\hbox{, }z=0)\left|
0,\varepsilon \right\rangle.
\end{equation}
Note that for $\lambda =1$, we have
\begin{equation}
U(\lambda =1\hbox{, }z=0)=I
\end{equation}
and the state $\left| \psi \right\rangle $ are nothing but the ground state $%
\left| 0,\varepsilon \right\rangle $ which is annihilated by the operator $%
A^{-}$ ($A^{-}\left| 0,\varepsilon \right\rangle =0$).

Note that the solution $\left| \psi \right\rangle $ is a linear combination
of the states $\left| 2k,\varepsilon \right\rangle $ ($k=0$, $1$, $2$, ...)
\begin{equation}
\left| \psi \right\rangle =\sum\limits_{k=0}^\infty c_{2k}\left|
2k,\varepsilon \right\rangle,
\end{equation}
where the coefficients $c_{2k}$ are given by
\begin{equation}
c_{2k}=\left( \frac{\left( \lambda -1\right) }{2\left( \lambda +1\right) }%
\right) ^k\frac 1{\Gamma (k+1)}\sqrt{\frac{\Gamma (2k+1)\Gamma \left(
1+k+\frac 1{3\varepsilon }\right) \Gamma (\frac 1{3\varepsilon }+\frac 32)}{%
\Gamma (\frac 1{3\varepsilon }+1)\Gamma (k+\frac 1{3\varepsilon }+\frac 32)}}%
e^{-i\alpha e_{2k}}c_0.
\end{equation}
The coefficients $c_0$ is the normalization constant which can be evaluated
by imposing the normalization condition \hspace{0.5cm}$\left\langle \psi
\right. \left| \psi \right\rangle =1$. We obtain
\begin{equation}
c_0=\left[ \sum\limits_{k=0}^\infty \left| \frac{\left( \lambda -1\right) }{%
2\left( \lambda +1\right) }\right| ^{2k}\frac{\Gamma (2k+1)\Gamma \left(
1+k+\frac 1{3\varepsilon }\right) \Gamma (\frac 1{3\varepsilon }+\frac 32)}{%
\left( \Gamma (k+1)\right) ^2\Gamma (\frac 1{3\varepsilon
}+1)\Gamma (k+\frac 1{3\varepsilon }+\frac 32)}\right] ^{-\frac
12}.
\end{equation}

\subsection{Analytical representation of Gazeau-Klauder coherent states}

It is well known that the analytical representation enable one to
find simpler solution of a number of problems, exploiting the
theory of analytical entire functions. In this section,
generalizing the pioneering work of Bargmann $\left[ 29\right] $
for the usual harmonic oscillator, we will study the analytical
representation of the extended Weyl-Heisenberg algebra (dynamical
algebra for the $x^4$-anharmonic oscillator). We recall that in
the anlytical representation of the standard harmonic oscillator,
the creation operator $a^{+}$ is the multiplication by $z$ while
the operator $a^{-}$ is the differentiation with respect to $z$

We define the analytic space as a space of functions which are holomorphic
on a ring $D$ of the complex plane. The scalar product is written with an
integral of the form
\begin{equation}
\left\langle f\right. \left| g\right\rangle =\int \overline{f(z)}\hbox{ }%
g(z)\hbox{ }d\mu (z),
\end{equation}
where $d\mu (z)$ is the measure defined above (see eq. $\left( 20\right) $).
Let $\left| f\right\rangle $ be an arbitrary state of the system under study
\begin{equation}
\left| f\right\rangle =\sum\limits_{n=0}^\infty f_n\left|
n,\varepsilon \right\rangle \hbox{\hspace{1cm}with\hspace{1cm}
}\sum\limits_{n=0}^\infty \left| f_n\right| ^2<\infty
\end{equation}
Following the construction of $\left[ 29\right] $, any state $\left|
f\right\rangle $ is represented, in the analytical representation, by a
function of the complex variable $z$ (using the coherent states associated
with $x^4$-anharmonic oscillator)
\begin{equation}
f(z)=\left\langle \overline{z},-\alpha \right. \left|
f\right\rangle =\sum\limits_{n=0}^\infty \frac{z^ne^{i\alpha
e_n}}{\sqrt{F(n)}}f_n,
\end{equation}
where the variable $z$ belong to the domain $D=\mathbf{C}$ of definition of
the eigenvalues of $A^{-}$ (annihilation operator). In particular to the
basis vectors $\left| n,\varepsilon \right\rangle $, there correspond the
monomials
\begin{equation}
\left\langle \overline{z},-\alpha \right. \left| n,\varepsilon
\right\rangle =\frac{z^ne^{i\alpha e_n}}{\sqrt{F(n)}}.
\end{equation}
Using the above considerations, we represent in the analytical
representation the annihilation operator $A^{-}$by
\begin{equation}
A^{-}=\left( 1+3\varepsilon \right) \frac d{dz}+\frac{3\varepsilon }2z\frac{%
d^2}{dz^2},
\end{equation}
the creation operator $A^{+}$%
\begin{equation}
A^{+}=z,
\end{equation}
and the operator number $N$ by
\begin{equation}
N=z\frac d{dz}.
\end{equation}
Remark that, in the limit $\varepsilon \rightarrow 0$, we have the
standard analytical representation of the Weyl-Heisenberg algebra:
\\ $A^{-}\equiv
a^{-}=\frac d{dz}$ , \hspace{0.5cm}$A^{+}\equiv a^{+}=z$, \hspace{0.5cm}$%
N=a^{+}a^{-}=z\frac d{dz}$.\\The analytical representation exists
if we have a measure such that
\begin{equation}
\int \left| z,\alpha \right\rangle \left\langle z,\alpha \right|
d\mu (z)=I_{\mathcal{H}}.
\end{equation}
The existence of the measure ensures that scalar product of the
representation takes the form $\left( \mathrm{110}\right) .$

\subsection{Analytical representation of Generalized Intelligent States for $%
\mathbf{x}^4$-anharmonic oscillator}

Let us now construct and discuss, using the analytical representation, the $%
x^4$-anharm-onic oscillator generalized intelligent states. We constructed
the $x^4$-oscillator equal variance $\left| z,\alpha \right\rangle $ ($%
\equiv \left| z,\lambda =1,\alpha \right\rangle $, the Gazeau-Klauder
coherent states or the eigenstates of the annihilation operator $A^{-}$ ).
These states form an overcomplete family of states (resolving the unity by
integration with respect to measure given by equation $\left( 24\right) $)
and provide a representation of any state $\left| \psi \right\rangle $ by an
entire function $\left\langle \overline{z},-\alpha \right. \left| \psi
\right\rangle =\psi (z)$. In what follows we will use the analytical
representation discussed previously to convert the eigenvalues equation $%
\left( 86\right) $ into homogenous differential equation. The latter permits
the construction of the generalized intelligent states, corresponding to the
system under consideration, in an analytical way. Indeed, using the analytic
realization of the creation $A^{+}$ and annihilation $A^{-}$ operators, the
equation (we denote for a while the eigenvalue by $z^{\prime }$)
\begin{equation}
\left( \left( 1+\lambda \right) A^{-}+\left( 1-\lambda \right)
A^{+}\right) \left| z^{\prime },\lambda ,\alpha \right\rangle
=2z^{\prime }\left| z^{\prime },\lambda ,\alpha \right\rangle,
\end{equation}
now reads
\begin{equation}
\left\{ \left( \left( 1+\lambda \right) \frac{3\varepsilon }2\right) \left[
\frac 2{3\varepsilon }\left( 1+3\varepsilon \right) \frac d{dz}+z\frac{d^2}{%
dz^2}\right] +\left( 1-\lambda \right) z\right\} \Phi _{\left(
z^{\prime },\lambda \right) }(z)=2z^{\prime }\Phi _{\left(
z^{\prime },\lambda \right) }(z).
\end{equation}
By means of simple substitutions the above equation is reduced to the Kummer
equation for the confluent hypergeometric function $_1F_1\left( a,b;z\right)
$ $\left[ 30\right] $. So that we have the following solution
\begin{equation}
\Phi _{\left( z^{\prime },\lambda \right) }(z)=\exp \left( cz\right) \hbox{ }%
_1F_1\left( a,b,-2cz\right),
\end{equation}
where
\begin{eqnarray}
a &=&1+\frac 1{3\varepsilon }-\frac{z^{\prime }}{2\mu c},%
\hspace{0.5cm}b=\frac 2{3\varepsilon }+2,\hspace{0.5cm}%
c^2=-\frac \upsilon \mu  \nonumber \\
\mu &=&\left( 1+\lambda \right) \frac{3\varepsilon }2,%
\hspace{1.5cm}\upsilon =\left( 1-\lambda \right).
\end{eqnarray}
Thus, we obtain the $x^4$-anharmonic oscillator intelligent states in the
coherent states representation in the form (up to the normalization
constant)
\begin{equation}
\left\langle z^{\prime },\lambda ,\alpha \right. \left| z,\alpha
\right\rangle =\exp \left( c^{*}z\right) \hbox{ }_1F_1\left(
a^{*},b,-2c^{*}z\right),
\end{equation}
where the parameters $a$, $b$ and $c$ are defined in the formulae $\left(
121\right) $. Using the power series of the confluent hypergeometric
function $_1F_1\left( a,b,z\right) $, we get at $\lambda =1$ the
Gazeau-Klauder coherent states discussed in the section 2 (up to
normalization constant)
\begin{equation}
\left\langle z^{\prime },\lambda =1,\alpha \right. \left| z,\alpha
\right\rangle =\hbox{ }_0F_1\left( 2+\frac 2{3\varepsilon },\frac
2{3\varepsilon }z\hbox{ }\overline{z^{\prime }}\right).
\end{equation}
We note also that the generalized intelligent states for the harmonic
oscillator can be obtained from the equation $\left( 122\right) $ in the
limit $\varepsilon \rightarrow 0$ (or from the differential equation $\left(
119\right) $ by setting $\varepsilon =0$). Thus, we have
\begin{equation}
\Phi _{\left( z^{\prime },\lambda \right) }(z)=\Phi _{\left(
z^{\prime },\lambda \right) }(0)\exp \left( \frac{2z^{\prime
}}{1+\lambda }z+\left( \frac{\lambda -1}{\lambda +1}\right)
\frac{z^2}2\right)  \label{ju}.
\end{equation}
A solution which coincides with the result obtained in the section 4 (see
the equation $\left( 97\right) $).

\section{Conclusion}

In this work, we have constructed the Generalized Intelligent (Squeezed $%
\left| \lambda \right| \neq 1$ and Coherent $\left| \lambda \right| =1$)
States (GIS) for the $x^4$-anharmonic oscillator. This construction is based
on the minimization of the Robertson-Schr\"odinger uncertainty relation. The
operators generating the (GIS) families, by their actions on the ground
state of the quantum system under consideration, are introduced. In the case
where the parameter $\lambda =1$, we recover the Gazeau-Klauder coherent
states corresponding to the $x^4$-anharmonic oscillator. Their properties
(continuity, temporal stability, overcompleteness and action identity) are
studied. We also shown that there is a dynamical algebra, generated by
raising and lowering operators, which may be seen as an extended
Weyl-Heisenberg algebra and is isomorphic to $su(1,1)$. The limit of the
standard harmonic oscillator is discussed. In order to get the Generalized
Intelligent States in an analytical way, we considered the analytical
representation. The advantage of this representation is clearly illustrated.
It should be noted that this work constitute another application of the
recent developments in the construction of the Generalized Intelligent
States obtained previously in $\left[ 28\right] $ where a first application
was made for the infinite square well and P\"oschl-Teller potentials. A
further extension concern the construction of the Perelomov coherent states
type for an arbitrary quantum system and compare them with Gazeau-Klauder
ones. This matter will be considered in a forthcoming work.

\begin{quote}
\textbf{Acknowledgements}

The authors are grateful to the referee for critical comments and extensive
suggestions on an earlier draft of the manuscript, which helped greatly to
improve the clarity and conciseness of the presentation.
\end{quote}

\newpage\

\end{document}